\documentclass[aps,prd,12pt,nofootinbib,notitlepage]{revtex4-1}

\usepackage{amsfonts}
\usepackage{amsmath,epsfig,bm}
\usepackage{graphicx}
\usepackage{bm}
\usepackage{color}

\usepackage{tabularx} 

\usepackage{times}

\newcommand{\be}{\begin{equation}}
\newcommand{\ee}{\end{equation}}
\newcommand{\ba}{\begin{eqnarray}}
\newcommand{\ea}{\end{eqnarray}}
\newcommand{\bd}{\begin{displaymath}}
\newcommand{\ed}{\end{displaymath}}

\def\oneth{{\textstyle{\frac{1}{3}}}}
\def\twoth{{\textstyle{\frac{2}{3}}}}

\def\rt3{\sqrt{3}}
\def\rt6{\sqrt{6}}
\def\mL2{(m_{\phi}L)^2}

\begin{document}

\title{{\bf Is Hyperon Polarization in Relativistic Heavy Ion Collisions Connected to Axial U(1) Symmetry Breaking at High Temperature?}}
\author{Joseph I. Kapusta$^1$, Ermal Rrapaj$^{1,2}$, and Serge Rudaz$^1$}
\affiliation{$^1$School of Physics and Astronomy, University of Minnesota, Minneapolis, Minnesota 55455, USA \\
$^2$Department of Physics, University of California, Berkeley, CA 94720, USA}

\vspace{.3cm}

\parindent=20pt

\begin{abstract}
Experiments at the Relativistic Heavy Ion Collider (RHIC) have measured the net polarization of $\Lambda$ and $\bar{\Lambda}$ hyperons and attributed it to a coupling between their spin and the vorticity of the fluid created in heavy ion collisions, but how the spin comes to equilibrium with vorticity is an open problem.  Recently we found that vorticity fluctuations and helicity flip of strange quarks in quark-gluon plasma through perturbative QCD processes resulted in equilibration times far too long to be relevant.  Here we consider the Nambu--Jona-Lasinio model with the inclusion of the six-quark Kobayashi--Maskawa--'t Hooft interaction which breaks axial U(1).  Using instanton inspired models for the temperature dependence of the axial symmetry breaking, we find that constituent strange quarks can reach spin equilibrium at temperatures below about 170 MeV, just before they hadronize to form hyperons.
\end{abstract}
\date{\today}

\maketitle


Quantum Chromodynamics (QCD) with three flavors of massless quarks has an 
SU(3)$_{\rm L} \times $ SU(3)$_{\rm R} \times $ U(1)$_{\rm V} \times $U(1)$_{\rm A}$ symmetry.  The left-right symmetry is explicitly broken by current quark masses.  When $m_u = m_d < m_s$ this reduces to SU(2)$_f \times $ U(1)$_{\rm V} \times $U(1)$_{\rm A}$.  The vector symmetry is associated with baryon number conservation.  If the up and down quark masses were zero, chiral symmetry would be restored in a second order transition at a critical temperature around 160 MeV.  When their masses are nonzero but small the transition is a rapid crossover.  Even before the formulation of QCD, it was suggested that there ought to be a 
six-quark U(1)$_{\rm A}$ symmetry breaking term in the effective action of determinantal form to solve the problem of the suprisingly large mass of the $\eta'$ meson \cite{KM1970,KKM1971}.  This interaction term is
\be
{\cal L}_6 = g_D \left[ \det\left({\bar q}_i (1 + \gamma_5) q_j \right) + \det\left({\bar q}_i (1 - \gamma_5) q_j \right) \right]
\label{L6}
\ee
where the determinant refers to flavor and where each matrix entry is a color singlet.  Incorporation of this term into the Nambu--Jona-Lasinio (NJL) model provides for a very good hadron phenomenology \cite{2Treview,Kunihiro4KM}.  Independently, it was argued that an unbroken U(1)$_{\rm A}$ symmetry would imply an isoscalar pseudoscalar boson with mass less than $\sqrt{3} m_{\pi}$, which has never been observed, a conundrum that became known as the U(1)$_{\rm A}$ problem \cite{WeinbergU1}.  In QCD it is known that there is an explicit quantum breaking of the U(1)$_{\rm A}$ symmetry resulting from the chiral anomaly 
\be
\partial_{\mu} J^{\mu}_{\rm A} = \frac{\alpha_s N_f }{4 \pi} F_a^{\mu\nu} \tilde{F}^a_{\mu\nu} +2 i m \bar{q} \gamma_5 q
\ee
where $J^{\mu}_{\rm A} = \bar{q} \gamma_{\mu} \gamma_5 q$.  The axial symmetry is thought to be partially, but never fully, restored at high temperature.  

Experiments at the Relativistic Heavy Ion Collider (RHIC) have provided an abundance of data on the hot, dense matter created in heavy ion collisions \cite{QMseries}.  Among these, the polarization of $\Lambda$ and $\bar{\Lambda}$ hyperons was proposed as an observable that provides information on collective flow, in particular vorticity \cite{Wang1,Becattini1}.  The vorticity arises in non-central heavy ion collisions where the produced matter has considerable angular momentum.  The spins of the $\Lambda$ and $\bar{\Lambda}$ couple to the vorticity, resulting in a splitting in energy between particles with spin parallel and antiparallel to the vorticity.  The decay products of these hyperons are used to infer their polarizations.  Measurements of the polarizations have been made by the STAR collaboration from the lowest to the highest beam energies at RHIC \cite{FirstSTAR,Nature,SecondSTAR}, noting that RHIC produces matter with the highest vorticity ever observed.  
The standard picture of $\Lambda$ and $\bar{\Lambda}$ polarization in non-central heavy ion collisions assumes equipartition of energy \cite{Becattini2,Becattini3}.  The spin-vorticity coupling is the same for baryons and antibaryons, which is approximately what is observed.  Within the quark model the spin of the $\Lambda$ is carried by the strange quark \cite{Jennings1,Cohen}.  One scenario posits that the $s$ and $\bar s$ quarks become polarized in the quark-gluon plasma phase and pass that polarization on to the $\Lambda$ and $\bar{\Lambda}$ during hadronization.  In Ref. \cite{us} we calculated the relaxation time for the strange quark spin to come to equilibrium with the vorticity via two mechanisms: vorticity fluctuations, and helicity flip in scatterings between strange quarks and light quarks and gluons using perturbative QCD.  With reasonable parameters both mechanisms lead to equilibration times orders of magnitude too large to be relevant to heavy ion collisions. 

The crossover between hadrons and quarks and gluons happens at temperatures in the neighborhood of 155 MeV.  Perturbation theory for QCD is not well behaved at such low temperatures.  This has led to various models to describe the strongly interacting quark-gluon plasma.  The interaction ${\cal L}_6$ is particularly intriquing in the context of hyperon polarization because it flips the helicities of the quarks.  For example, an incoming right-handed s-quark emerges as a left-handed s-quark.  This motivates us to study helicity flip rates in the NJL model\footnote{It should be noted that the NJL model does not provide for color confinement.} with the incorporation of  ${\cal L}_6$.   

Exactly this six-quark effective interaction arises from instanton physics.  In Euclidean space with volume $\beta V$ the instanton contribution to the partition function in the dilute gas approximation, including one-loop quantum corrections, is \cite{tHooft1}
\be
\ln \, Z_{\rm DGA} = 2 \beta V \int_0^{\infty} d\rho \, d(\rho)
\ee
This involves an integration over the instanton size $\rho$.  The density of instantons is defined by
\be
d(\rho) \equiv \frac{C(N_c)}{\rho^5}
\left( \frac{4\pi^2}{g^2}\right)^{2N_c} \exp\left(-\frac{8\pi^2}{\bar{g}^2}\right) 
\ee
The factor of 2 includes the contribution of anti-instantons.  The $C(N_c)$ is a group-theoretic factor.  In the Pauli-Villars 
regularization scheme
\be
C(N_c) = \frac{0.260156}{(N_c-1)!(N_c-2)!} \, \xi^{-(N_c-2)}
\ee
with $\xi = 1.33876$.  Quantum fluctuations amount to replacing the coupling constant $g^2$ with the renormalization group running coupling
\be
\bar{g}^2 = \frac{8\pi^2}{b \ln (1/\rho\Lambda_{\rm R})} \,, \;\;\;\;\; b = \oneth (11 N_c - 2 N_f )
\ee
in the exponential factor, although this replacement is presumed to happen (at the next order) in the pre-exponential factor as well.  Here $\Lambda_{\rm R}$ 
is the QCD scale parameter in the Pauli-Villars scheme.  
When light quarks are included, not only is the running coupling affected but there is also a factor of $\xi m_f \rho$ multiplying the instanton density for each flavor.  This suppresses the instanton density but also makes the ultraviolet divergence worse (in the vacuum).  However, it was soon shown that the current quark mass in this factor should be replaced by $m_f - \twoth \pi^2 \langle \bar{q}_f q_f \rangle \rho^2$ because the quark condensate $\langle \bar{q}_f q_f \rangle$ does not vanish in the physical vacuum \cite{Shifman1980}.  Both $m_f$ and $\langle \bar{q}_f q_f \rangle$ are to be evaluated at the renormalization scale $\rho^{-1}$.  Note that $\langle \bar{q}_f q_f \rangle$ is negative.

One needs to address the divergence of the integration over instanton size.  It was suggested in Ref. \cite{Diakonov1984} that this can be done approximately by the factor
\bd
\exp\left[ - \left(\frac{b-4}{2}\right) \frac{\rho^2}{{\bar \rho}^2} \right]
\ed
to take into account repulsive instanton and anti-instanton interactions.  Here $\bar{\rho} \approx 0.33$ fm \cite{Schafer_review}.  There are various other phenomenological models which we shall not go into here.  

At high temperature, color electric fields are screened just like in QED plasma.  The temperature should provide an infrared cutoff on instanton sizes.  It is necessary to compute the one-loop quantum correction in the background field of an instanton or anti-instanton at finite temperature.  This is a formidable task, but has been done at finite temperature \cite{PY1980,GPY} and with chemical potentials \cite{Carvalho,Baluni,Shuryakmu,Abrikosov}.  The result is that the instanton density $d(\rho)$ at large $\rho$ is multiplied by a cutoff factor
\be
\exp\left[ - \frac{2\pi^2}{g^2} m_{\rm el}^2 \, \rho^2\right]
\label{Tmucut}
\ee
where the color electric screening mass is
\be
m_{\rm el}^2 = g^2 \left[ \left( \frac{N_c}{3} +\frac{N_f}{6} \right)T^2 + \frac{1}{2\pi^2} \sum_f \mu_f^2 \right]
\ee
The modification is minor at intermediate values of $\rho$ and vanishes as $\rho \rightarrow 0$.  This means that at nonzero $T$ and $\mu_f$ the $\rho$ integration is both infrared and ultraviolet convergent.

Now we come to the six-quark effective interaction arising specifically from instantons.  It is \cite{tHooft1,Shifman1980}
\be
{\cal L}_6 = \int_0^{\infty} d\rho \, d(\rho) \left\{ \prod_f \left[ - \frac{\pi^2 \xi}{N_c} \, \bar{q}_f (1 + \gamma_5 ) q_f \, \rho^3 \right] 
+ (\gamma_5 \rightarrow - \gamma_5) + \cdot\cdot\cdot \right\} 
\ee
This neglects the current quark masses, which numerically is legitimate for the up and down quarks.  The strange quark mass is irrelevant for the interactions we are concerned with below.  The dots refer to extra contributions arising from Fierz transformations.  These involve color currents which are usually ignored in phenomenological applications of the extended NJL model and so we drop them as well.

In order to study elastic helicity-flip reactions we apply a mean field approximation to the six-quark effective interaction to reduce it to an effective four-quark interaction.  It is  
\ba
{\cal L}_{6 \rightarrow 4} = g_D &\Big\{ & \langle {\bar u} u \rangle \left[ {\bar d} (1 + \gamma_5) d \times {\bar s} (1 + \gamma_5) s 
+ {\bar d} (1 - \gamma_5) d \times {\bar s} (1 - \gamma_5) s \right] \nonumber \\
&+& \, \langle {\bar d} d \rangle \left[ {\bar u} (1 + \gamma_5) u \times {\bar s} (1 + \gamma_5) s 
+ {\bar u} (1 - \gamma_5) u \times {\bar s} (1 - \gamma_5) s \right] \nonumber \\
&+& \, \langle {\bar s} s \rangle \left[ {\bar u} (1 + \gamma_5) u \times {\bar d} (1 + \gamma_5) d 
+ {\bar u} (1 - \gamma_5) u \times {\bar d} (1 - \gamma_5) d \right] \Big\}
\label{L6to4} 
\ea
with the recognition that $\langle \bar{q}_f \gamma_5 q_f \rangle = 0$ in the vacuum.  Upon reduction one more time it contributes to the effective (constituent) quark masses.
\be
{\cal L}_{6 \rightarrow 2} = 2 g_D \left[ \langle \bar{u} u \rangle \langle \bar{d} d \rangle \bar{s} s
+ \langle \bar{u} u \rangle \langle \bar{s} s \rangle \bar{d} d + \langle \bar{d} d \rangle \langle \bar{s} s \rangle \bar{u} u \right]
\ee
In what follows we will assume that $\langle \bar{u} u \rangle = \langle \bar{d} d \rangle$.

We now have a theoretically and phenomenologically motivated four-quark interaction among constituent, as opposed to current, quarks.  A left-handed s-quark can scatter from a left-handed u or d-quark to become a right-handed s-quark.  To make quantitative estimates we need numerical values for the parameters of the model.  A fit to the $\eta'$ mass in Refs. \cite{2Treview,Kunihiro4KM} results in the numerical value $g_D = -9.288/\Lambda^5$, where $\Lambda = 631.4$ MeV is a 3-momentum cut-off used in this non-renormalizable model.  In the same fit were the current quark masses of $m_u = m_d = 5.5$ MeV and $m_s = 135.7$ MeV, and the light quark condensate 
$\langle \bar{u} u \rangle = - (245 \, {\rm MeV})^3$.  Reference \cite{Fuku1} obtained $g_D = -11.32/\Lambda^5$ with the same cut-off and current quark masses by fitting the topological susceptibility as calculated with lattice QCD.  These lead to $g_D = -92.55$ and $-112.8$ GeV$^{-5}$, respectively.  We will use $g_D = -100$ GeV$^{-5}$, acknowledging a 10\% uncertainty.

The temperature dependence of the light quark condensate $\langle \bar{u} u \rangle$ is taken from Fig. 6.1 (case II) of Ref. \cite{2Treview}.  We also need the temperature dependence of the effective/constituent quark masses.  From Ref. \cite{2Treview} their $T=0$ values are 335 and 527 MeV for light and strange quarks, respectively.  Their temperature dependence is shown in Fig. 6.3 (case II) of the same reference.  Lattice QCD does not inform us on these masses.  The temperature dependence of $g_D$ has not been determined with any accuracy.

We may construct an instanton inspired model to estimate the temperature dependence of $g_D$.  Reference \cite{Kunihiro_PLB1988} suggested that it should be
\be
g_D(T) = g_D(0) \exp\left( - T^2/T_0^2\right)
\ee
This is based on evaluating Eq. (\ref{Tmucut}) at ${\bar \rho}$.  That work used $T_0 = 100$ MeV when $N_c = N_f = 3$.  We shall refer to this as case I.  On the other hand, following the suggestion of \cite{Diakonov1984} mentioned above, and neglecting logarithmic corrections, leads to
\be
g_D(T) \propto \int_0^{\infty} d\rho \, \rho^{b + 3N_f - 5} \exp \left[ - \left(\frac{b-4}{2}\right) \frac{\rho^2}{{\bar \rho}^2} \right]
\exp \left[ - \frac{1}{3} (2N_c + N_f) \left( \pi T\right)^2 \rho^2 \right]
\ee
With $N_c = N_f = 3$ this is
\be
g_D(T) = \frac{g_D(0)}{\left( 1 + 1.2 \pi^2 {\bar \rho}^2 T^2 \right)^7}
\ee
We shall refer to this as case II.  Both have a very strong temperature dependence.

Consider the reaction $a+b \rightarrow c+d$ where all particles are fermions.  The relaxation time $\tau(E)$ for species $a$ with energy $E$ as measured in the rest frame of the plasma is given by \cite{ChakrabortyKapusta2011,AlbrightKapusta2016,us}
\be
\frac{1 - f^{\rm eq}(E)}{\tau(E)}  = \frac{ {\cal N} \, T}{32 (2\pi)^3 E^2} \int \frac{ds}{s} \ln \left( 1 + e^{-s/4ET} \right) \int dt \, |{\cal M}(s,t)|^2 \,.
\label{tau}
\ee
Here ${\cal M}$ is the dimensionless amplitude for the reaction.  The ${\cal N}$ is a degeneracy factor for spin, color, and any other internal degrees of freedom.  Its value depends on how these variables are summed or averaged over in $|{\cal M}|^2$.  In order to obtain analytical results we dropped the Pauli suppression factors in the final state, which is a small corrrection at high temperature.

First consider the reaction $s+u \rightarrow s+u$ arising from the Lagrangian (\ref{L6to4}).  The amplitude for strange quark helicity flip is denoted by ${\cal M}(\sigma_s, \sigma_u; -\sigma_s, \sigma_u')$ where $\sigma_s$ is the helicity of the incoming strange quark, $\sigma_u$ is the helicity of the incoming up quark, and $\sigma_u'$ is the helicity of the outging up quark.  Using the method of Ref. \cite{Fearing} one readily finds that
\ba
\hat{{\cal M}}(+, +; -, -) & = & 2 \left( s - M_s^2 - M_u^2 \right) (1 - \cos\theta) \nonumber \\
\hat{{\cal M}}(+, +; -, +) & = & \hat{{\cal M}}(+, -; -, -) = \frac{2 M_u \left( s + M_s^2 - M_u^2 \right)}{\sqrt{s}} \sin\theta \nonumber \\
\hat{{\cal M}}(+, -; -, +) & = & \frac{2}{s} \left[ (M_s^2 + M_u^2) s - (M_s^2 - M_u^2)^2 \right](1 - \cos\theta) 
\ea
The hat means that there is a common overall factor of $g_D \langle \bar{d} d \rangle$.  The $M_s$ and $M_u$ are the constituent quark masses.  All of these amplitudes vanish when the scattering angle $\theta$ in the center-of-momentum frame is zero.  Due to the nature of the Kobayashi--Masakawa--'t Hooft interaction, a massless up quark is forced to flip its helicity when the strange quark does.  Using $\sqrt{s} = \sqrt{M_s^2 + p_*^2} + \sqrt{M_u^2 + p_*^2}$ and $t = -2p_*^2 ( 1 - \cos\theta)$ the integration over $t$ is easily done with the result
\ba
\int_{-4p_*^2}^0 dt \, |{\cal M}|^2(+, +; -, -) & = & \frac{16}{3 s} \left( s - M_s^2 - M_u^2 \right)^2 
\lambda(s,M_s^2,M_u^2) \, g_D^2 \, \langle \bar{d} d \rangle^2 \nonumber \\
\int_{-4p_*^2}^0 dt \, |{\cal M}|^2(+, +; -, +) & = & \frac{8 M_u^2}{3 s^2} \left( s + M_s^2 - M_u^2 \right)^2 
\lambda(s,M_s^2,M_u^2) \, g_D^2 \, \langle \bar{d} d \rangle^2 \nonumber \\
\int_{-4p_*^2}^0 dt \, |{\cal M}|^2(+, -; -, +) & = & \frac{16(M_s^2 + M_u^2)^2}{3 s^3} \left( s - M_s^2 - M_u^2 \right)^2 
\lambda(s,M_s^2,M_u^2) \, g_D^2 \, \langle \bar{d} d \rangle^2
\ea
Here $\lambda(s,M_s^2,M_u^2) = (s - M_s^2 - M_u^2)^2 - 4 M_s^2 M_u^2 = 4 s p_*^2$.  The above expressions are inserted into Eq. (\ref{tau}).  
We set ${\cal N} = 12$ to take into account scattering from both flavors of light quarks and anti-quarks, each of which comes in three colors.
 
\begin{figure}[hh]
\centering
\includegraphics[scale=0.6]{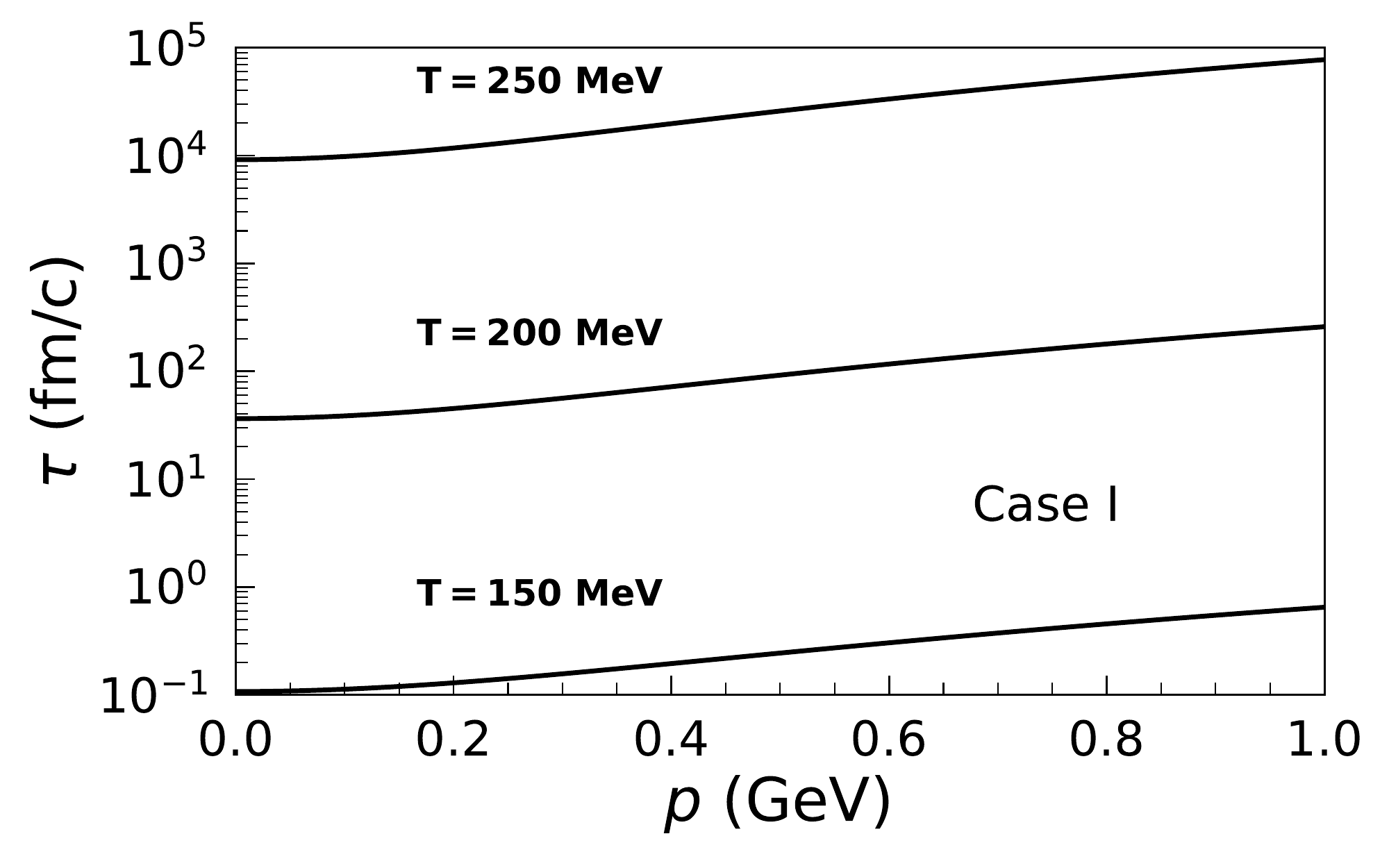}
\caption{Equilibration time for strange quark helicity as a function of momentum for three values of the temperature for case I modeling of the six-quark coupling $g_D(T)$.}
\label{caseI}
\end{figure}

\begin{figure}[hh]
\centering
\includegraphics[scale=0.6]{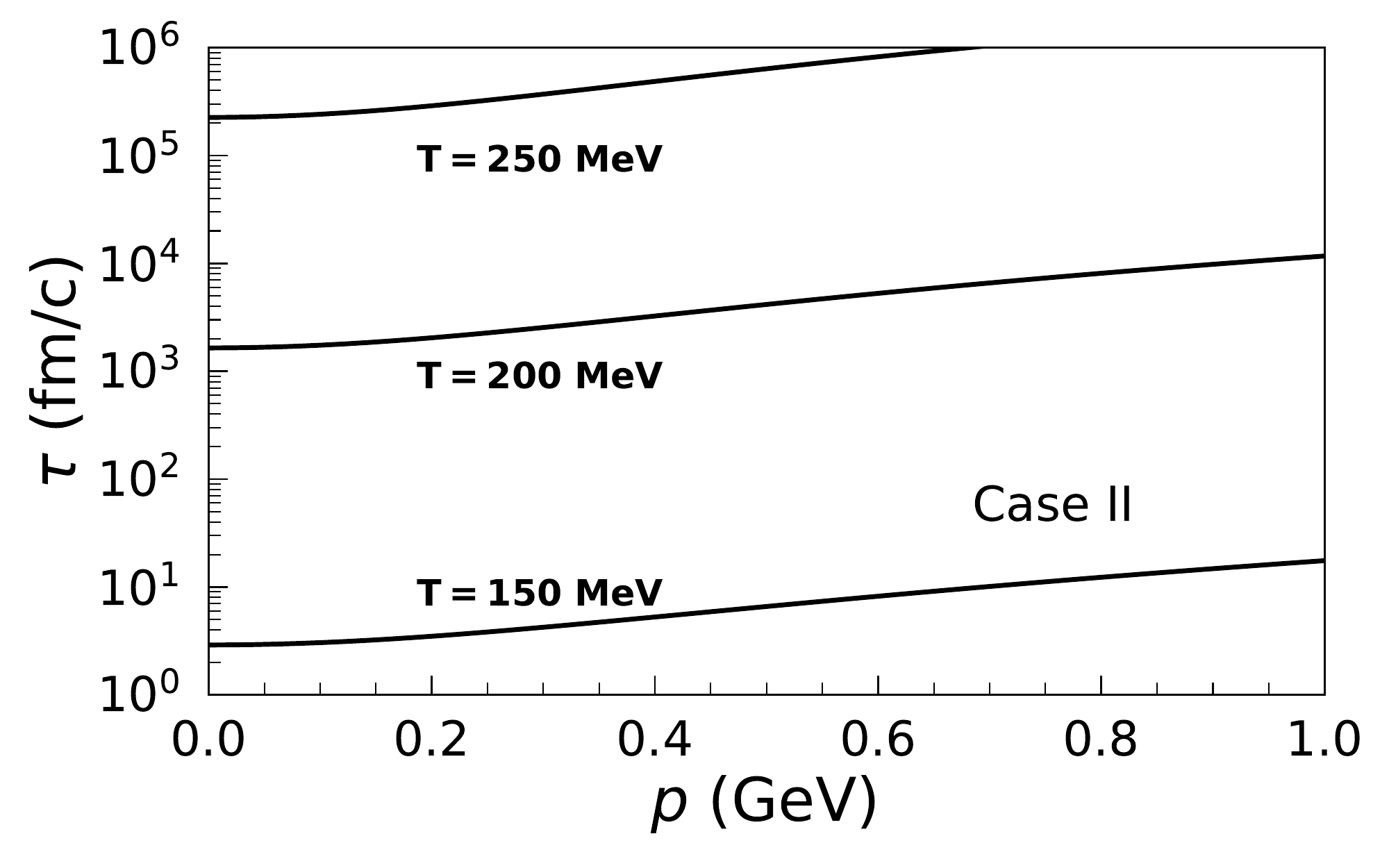}
\caption{Equilibration time for strange quark helicity as a function of momentum for three values of the temperature for case II modeling of the six-quark coupling $g_D(T)$.}
\label{caseII}
\end{figure}

\begin{figure}[hh]
\centering
\includegraphics[scale=0.6]{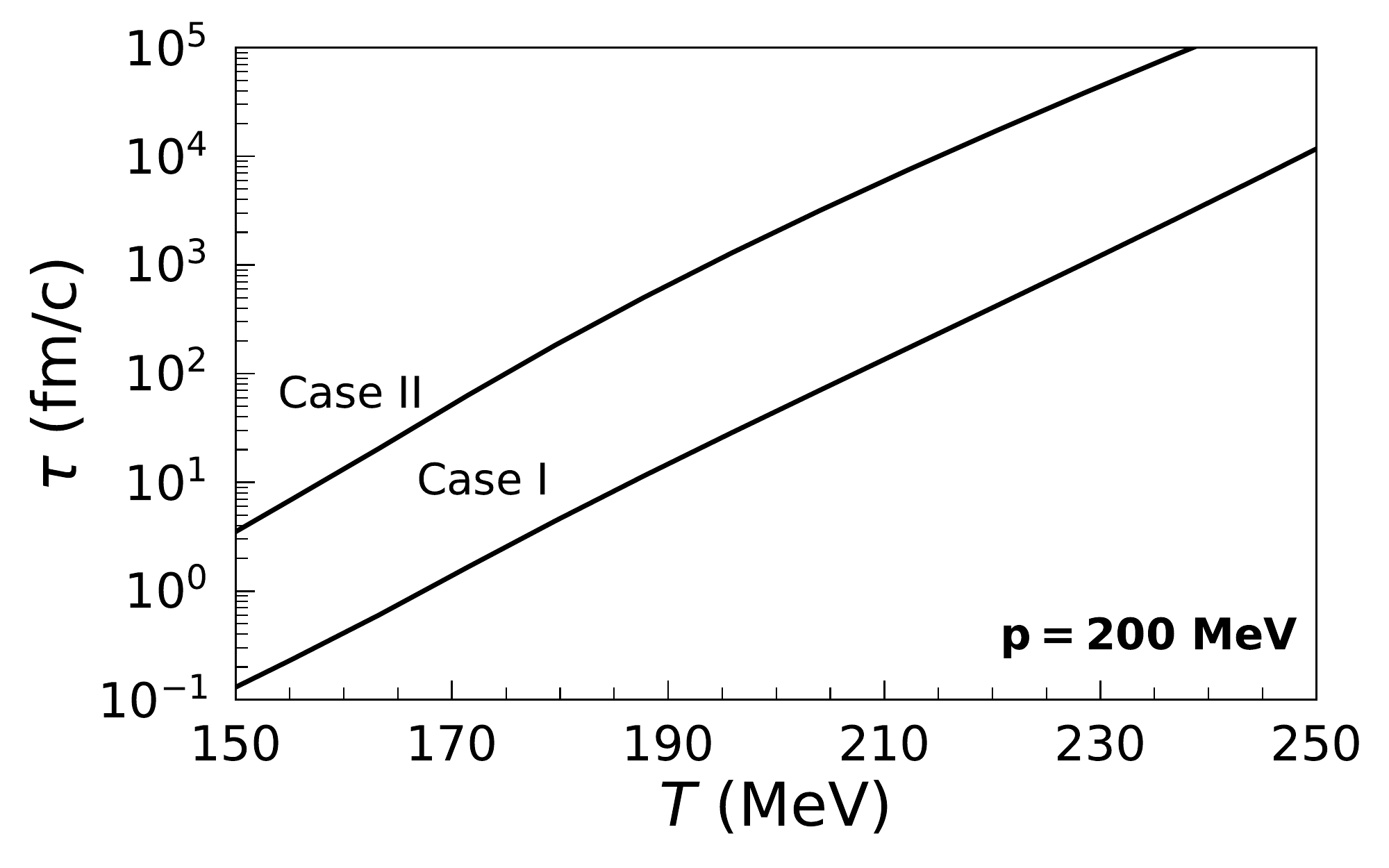}
\caption{Equilibration time for strange quark helicity as a function of temperature at a representative momentum for cases I and II modeling of the six-quark coupling $g_D(T)$.}
\label{compare}
\end{figure}

Figures \ref{caseI} and \ref{caseII} show the helicity flip equilibration time for strange quarks as a function of momentum for cases I and II, respectively.  There are two points to note.  First, $\tau$ is smallest at $p = 0$ and increases monotonically with $p$, being about an order of magnitude larger at $p = 1$ GeV.  Second, and more importantly, $\tau$ decreases dramatically with a decrease in temperature.  Figure \ref{compare} shows $\tau$ versus $T$ for a representative value of momentum $p = 200$ MeV.  There is about a factor 20 difference between the two models of $g_D(T)$.  This indicates the sensitivity to the product 
$g_D \langle \bar{u} u \rangle$, for which the rate is proportional to its square.  Nevertheless, due to the approximately exponential dependence of $\tau$ on $T$, case I has $\tau < 2$ fm/c when $T < 175$ MeV while case II has $\tau < 2$ fm/c when $T < 165$ MeV.  Since hadronization occurs at time scales on the order of 3-5 fm/c, or even a little longer, it means that strange quark helicities could come into equilibrium with its thermal and vortical environment.  The strange quarks and anti-quarks could then pass along their spin to the hyperons.

In principle one should use a density matrix to determine the spin relaxation time along the vorticity axis.  We have used the helicity flip has as a proxy, recognizing that it should be a very close estimate.  The reason is that the energy difference between spin parallel and anti-parallel to the vorticity is less than 10 MeV, which is only a few percent of the temperature.  In that case the difference between helicity and vorticity equilibration times should be negligible.

\newpage

The results of the calculations in this paper, combined with those in Ref. \cite{us}, thus point to the following scenario.  Strange and anti-strange quarks may or may not be produced with an initial global polarization.  Vorticity fluctuations and helicity flip scattering as calculated from perturbative QCD at finite temperature would not affect the polarization as the quark-gluon plasma evolves.  However, as the expanding matter approaches the transition to hadrons, the matter becomes strongly interacting due to non-perturbative effects and constituent $s$ and $\bar{s}$ quarks achieve spin and helicity equilibration with the vorticity.  The resulting polarization is passed on to the $\Lambda$ and $\bar{\Lambda}$ hyperons, whose decays are them measured by the experiments.

In summary, we have proposed a connection between axial U(1) symmetry breaking and it's temperature dependence and experimental measurements at RHIC.  The theoretical status of axial U(1) symmetry restoration with increasing temperature is highly uncertain; see \cite{latest} for the latest overview.  We hope that our work provokes further investigation of this important topic.  

\section*{Acknowledgement}
The work of JIK was supported by the U.S. Department of Energy Grant DE-FG02-87ER40328.  The work of ER was supported by the U.S. National Science Foundation Grant PHY-1630782 and by the Heising-Simons Foundation Grant 2017-228.


\end{document}